\documentclass[]{article}
\usepackage{cmap} 
\usepackage{ifthen}
\usepackage[T1]{fontenc}
\usepackage[utf8]{inputenc}
\usepackage{arxiv}
\usepackage{scipy}
\makeatletter
\def\PY@reset{\let\PY@it=\relax \let\PY@bf=\relax%
    \let\PY@ul=\relax \let\PY@tc=\relax%
    \let\PY@bc=\relax \let\PY@ff=\relax}
\def\PY@tok#1{\csname PY@tok@#1\endcsname}
\def\PY@toks#1+{\ifx\relax#1\empty\else%
    \PY@tok{#1}\expandafter\PY@toks\fi}
\def\PY@do#1{\PY@bc{\PY@tc{\PY@ul{%
    \PY@it{\PY@bf{\PY@ff{#1}}}}}}}
\def\PY#1#2{\PY@reset\PY@toks#1+\relax+\PY@do{#2}}

\expandafter\def\csname PY@tok@w\endcsname{\def\PY@tc##1{\textcolor[rgb]{0.73,0.73,0.73}{##1}}}
\expandafter\def\csname PY@tok@c\endcsname{\let\PY@it=\textit\def\PY@tc##1{\textcolor[rgb]{0.25,0.50,0.56}{##1}}}
\expandafter\def\csname PY@tok@cp\endcsname{\def\PY@tc##1{\textcolor[rgb]{0.00,0.44,0.13}{##1}}}
\expandafter\def\csname PY@tok@cs\endcsname{\def\PY@tc##1{\textcolor[rgb]{0.25,0.50,0.56}{##1}}\def\PY@bc##1{\setlength{\fboxsep}{0pt}\colorbox[rgb]{1.00,0.94,0.94}{\strut ##1}}}
\expandafter\def\csname PY@tok@k\endcsname{\let\PY@bf=\textbf\def\PY@tc##1{\textcolor[rgb]{0.00,0.44,0.13}{##1}}}
\expandafter\def\csname PY@tok@kp\endcsname{\def\PY@tc##1{\textcolor[rgb]{0.00,0.44,0.13}{##1}}}
\expandafter\def\csname PY@tok@kt\endcsname{\def\PY@tc##1{\textcolor[rgb]{0.56,0.13,0.00}{##1}}}
\expandafter\def\csname PY@tok@o\endcsname{\def\PY@tc##1{\textcolor[rgb]{0.40,0.40,0.40}{##1}}}
\expandafter\def\csname PY@tok@ow\endcsname{\let\PY@bf=\textbf\def\PY@tc##1{\textcolor[rgb]{0.00,0.44,0.13}{##1}}}
\expandafter\def\csname PY@tok@nb\endcsname{\def\PY@tc##1{\textcolor[rgb]{0.00,0.44,0.13}{##1}}}
\expandafter\def\csname PY@tok@nf\endcsname{\def\PY@tc##1{\textcolor[rgb]{0.02,0.16,0.49}{##1}}}
\expandafter\def\csname PY@tok@nc\endcsname{\let\PY@bf=\textbf\def\PY@tc##1{\textcolor[rgb]{0.05,0.52,0.71}{##1}}}
\expandafter\def\csname PY@tok@nn\endcsname{\let\PY@bf=\textbf\def\PY@tc##1{\textcolor[rgb]{0.05,0.52,0.71}{##1}}}
\expandafter\def\csname PY@tok@ne\endcsname{\def\PY@tc##1{\textcolor[rgb]{0.00,0.44,0.13}{##1}}}
\expandafter\def\csname PY@tok@nv\endcsname{\def\PY@tc##1{\textcolor[rgb]{0.73,0.38,0.84}{##1}}}
\expandafter\def\csname PY@tok@no\endcsname{\def\PY@tc##1{\textcolor[rgb]{0.38,0.68,0.84}{##1}}}
\expandafter\def\csname PY@tok@nl\endcsname{\let\PY@bf=\textbf\def\PY@tc##1{\textcolor[rgb]{0.00,0.13,0.44}{##1}}}
\expandafter\def\csname PY@tok@ni\endcsname{\let\PY@bf=\textbf\def\PY@tc##1{\textcolor[rgb]{0.84,0.33,0.22}{##1}}}
\expandafter\def\csname PY@tok@na\endcsname{\def\PY@tc##1{\textcolor[rgb]{0.25,0.44,0.63}{##1}}}
\expandafter\def\csname PY@tok@nt\endcsname{\let\PY@bf=\textbf\def\PY@tc##1{\textcolor[rgb]{0.02,0.16,0.45}{##1}}}
\expandafter\def\csname PY@tok@nd\endcsname{\let\PY@bf=\textbf\def\PY@tc##1{\textcolor[rgb]{0.33,0.33,0.33}{##1}}}
\expandafter\def\csname PY@tok@s\endcsname{\def\PY@tc##1{\textcolor[rgb]{0.25,0.44,0.63}{##1}}}
\expandafter\def\csname PY@tok@sd\endcsname{\let\PY@it=\textit\def\PY@tc##1{\textcolor[rgb]{0.25,0.44,0.63}{##1}}}
\expandafter\def\csname PY@tok@si\endcsname{\let\PY@it=\textit\def\PY@tc##1{\textcolor[rgb]{0.44,0.63,0.82}{##1}}}
\expandafter\def\csname PY@tok@se\endcsname{\let\PY@bf=\textbf\def\PY@tc##1{\textcolor[rgb]{0.25,0.44,0.63}{##1}}}
\expandafter\def\csname PY@tok@sr\endcsname{\def\PY@tc##1{\textcolor[rgb]{0.14,0.33,0.53}{##1}}}
\expandafter\def\csname PY@tok@ss\endcsname{\def\PY@tc##1{\textcolor[rgb]{0.32,0.47,0.09}{##1}}}
\expandafter\def\csname PY@tok@sx\endcsname{\def\PY@tc##1{\textcolor[rgb]{0.78,0.36,0.04}{##1}}}
\expandafter\def\csname PY@tok@m\endcsname{\def\PY@tc##1{\textcolor[rgb]{0.13,0.50,0.31}{##1}}}
\expandafter\def\csname PY@tok@gh\endcsname{\let\PY@bf=\textbf\def\PY@tc##1{\textcolor[rgb]{0.00,0.00,0.50}{##1}}}
\expandafter\def\csname PY@tok@gu\endcsname{\let\PY@bf=\textbf\def\PY@tc##1{\textcolor[rgb]{0.50,0.00,0.50}{##1}}}
\expandafter\def\csname PY@tok@gd\endcsname{\def\PY@tc##1{\textcolor[rgb]{0.63,0.00,0.00}{##1}}}
\expandafter\def\csname PY@tok@gi\endcsname{\def\PY@tc##1{\textcolor[rgb]{0.00,0.63,0.00}{##1}}}
\expandafter\def\csname PY@tok@gr\endcsname{\def\PY@tc##1{\textcolor[rgb]{1.00,0.00,0.00}{##1}}}
\expandafter\def\csname PY@tok@ge\endcsname{\let\PY@it=\textit}
\expandafter\def\csname PY@tok@gs\endcsname{\let\PY@bf=\textbf}
\expandafter\def\csname PY@tok@gp\endcsname{\let\PY@bf=\textbf\def\PY@tc##1{\textcolor[rgb]{0.78,0.36,0.04}{##1}}}
\expandafter\def\csname PY@tok@go\endcsname{\def\PY@tc##1{\textcolor[rgb]{0.20,0.20,0.20}{##1}}}
\expandafter\def\csname PY@tok@gt\endcsname{\def\PY@tc##1{\textcolor[rgb]{0.00,0.27,0.87}{##1}}}
\expandafter\def\csname PY@tok@err\endcsname{\def\PY@bc##1{\setlength{\fboxsep}{0pt}\fcolorbox[rgb]{1.00,0.00,0.00}{1,1,1}{\strut ##1}}}
\expandafter\def\csname PY@tok@kc\endcsname{\let\PY@bf=\textbf\def\PY@tc##1{\textcolor[rgb]{0.00,0.44,0.13}{##1}}}
\expandafter\def\csname PY@tok@kd\endcsname{\let\PY@bf=\textbf\def\PY@tc##1{\textcolor[rgb]{0.00,0.44,0.13}{##1}}}
\expandafter\def\csname PY@tok@kn\endcsname{\let\PY@bf=\textbf\def\PY@tc##1{\textcolor[rgb]{0.00,0.44,0.13}{##1}}}
\expandafter\def\csname PY@tok@kr\endcsname{\let\PY@bf=\textbf\def\PY@tc##1{\textcolor[rgb]{0.00,0.44,0.13}{##1}}}
\expandafter\def\csname PY@tok@bp\endcsname{\def\PY@tc##1{\textcolor[rgb]{0.00,0.44,0.13}{##1}}}
\expandafter\def\csname PY@tok@fm\endcsname{\def\PY@tc##1{\textcolor[rgb]{0.02,0.16,0.49}{##1}}}
\expandafter\def\csname PY@tok@vc\endcsname{\def\PY@tc##1{\textcolor[rgb]{0.73,0.38,0.84}{##1}}}
\expandafter\def\csname PY@tok@vg\endcsname{\def\PY@tc##1{\textcolor[rgb]{0.73,0.38,0.84}{##1}}}
\expandafter\def\csname PY@tok@vi\endcsname{\def\PY@tc##1{\textcolor[rgb]{0.73,0.38,0.84}{##1}}}
\expandafter\def\csname PY@tok@vm\endcsname{\def\PY@tc##1{\textcolor[rgb]{0.73,0.38,0.84}{##1}}}
\expandafter\def\csname PY@tok@sa\endcsname{\def\PY@tc##1{\textcolor[rgb]{0.25,0.44,0.63}{##1}}}
\expandafter\def\csname PY@tok@sb\endcsname{\def\PY@tc##1{\textcolor[rgb]{0.25,0.44,0.63}{##1}}}
\expandafter\def\csname PY@tok@sc\endcsname{\def\PY@tc##1{\textcolor[rgb]{0.25,0.44,0.63}{##1}}}
\expandafter\def\csname PY@tok@dl\endcsname{\def\PY@tc##1{\textcolor[rgb]{0.25,0.44,0.63}{##1}}}
\expandafter\def\csname PY@tok@s2\endcsname{\def\PY@tc##1{\textcolor[rgb]{0.25,0.44,0.63}{##1}}}
\expandafter\def\csname PY@tok@sh\endcsname{\def\PY@tc##1{\textcolor[rgb]{0.25,0.44,0.63}{##1}}}
\expandafter\def\csname PY@tok@s1\endcsname{\def\PY@tc##1{\textcolor[rgb]{0.25,0.44,0.63}{##1}}}
\expandafter\def\csname PY@tok@mb\endcsname{\def\PY@tc##1{\textcolor[rgb]{0.13,0.50,0.31}{##1}}}
\expandafter\def\csname PY@tok@mf\endcsname{\def\PY@tc##1{\textcolor[rgb]{0.13,0.50,0.31}{##1}}}
\expandafter\def\csname PY@tok@mh\endcsname{\def\PY@tc##1{\textcolor[rgb]{0.13,0.50,0.31}{##1}}}
\expandafter\def\csname PY@tok@mi\endcsname{\def\PY@tc##1{\textcolor[rgb]{0.13,0.50,0.31}{##1}}}
\expandafter\def\csname PY@tok@il\endcsname{\def\PY@tc##1{\textcolor[rgb]{0.13,0.50,0.31}{##1}}}
\expandafter\def\csname PY@tok@mo\endcsname{\def\PY@tc##1{\textcolor[rgb]{0.13,0.50,0.31}{##1}}}
\expandafter\def\csname PY@tok@ch\endcsname{\let\PY@it=\textit\def\PY@tc##1{\textcolor[rgb]{0.25,0.50,0.56}{##1}}}
\expandafter\def\csname PY@tok@cm\endcsname{\let\PY@it=\textit\def\PY@tc##1{\textcolor[rgb]{0.25,0.50,0.56}{##1}}}
\expandafter\def\csname PY@tok@cpf\endcsname{\let\PY@it=\textit\def\PY@tc##1{\textcolor[rgb]{0.25,0.50,0.56}{##1}}}
\expandafter\def\csname PY@tok@c1\endcsname{\let\PY@it=\textit\def\PY@tc##1{\textcolor[rgb]{0.25,0.50,0.56}{##1}}}

\def\PYZbs{\char`\\}
\def\PYZus{\char`\_}
\def\PYZob{\char`\{}
\def\PYZcb{\char`\}}

\def\PYZlt{\char`\<}
\def\PYZgt{\char`\>}
\def\PYZsh{\char`\#}
\def\PYZpc{\char`\%}

\def\PYZhy{\char`\-}
\def\PYZsq{\char`\'}
\def\PYZdq{\char`\"}

\makeatother

\providecommand*{\DUprovidelength}[2]{
  \ifthenelse{\isundefined{#1}}{\newlength{#1}\setlength{#1}{#2}}{}
}

\providecommand*{\DUrole}[2]{%
  \ifcsname docutilsrole#1\endcsname%
    \csname docutilsrole#1\endcsname{#2}%
  \else
    \csname DUrole#1\endcsname{#2}%
  \fi%
}

\DUprovidelength{\DUlineblockindent}{2.5em}
\ifthenelse{\isundefined{\DUlineblock}}{
  \newenvironment{DUlineblock}[1]{%
    \list{}{\setlength{\partopsep}{\parskip}
            \addtolength{\partopsep}{\baselineskip}
            \setlength{\topsep}{0pt}
            \setlength{\itemsep}{0.15\baselineskip}
            \setlength{\parsep}{0pt}
            \setlength{\leftmargin}{#1}}
    \raggedright
  }
  {\endlist}
}{}

\providecommand*{\DUroletitlereference}[1]{\textsl{#1}}

\ifthenelse{\isundefined{\hypersetup}}{
  \usepackage[colorlinks=true,linkcolor=blue,urlcolor=blue]{hyperref}
  \usepackage{bookmark}
  \urlstyle{same} 
}{}

\begin{document}
\title{Network visualizations with Pyvis and VisJS}\author{Giancarlo Perrone\\
Gary and Mary West Health Institute\\
La Jolla, CA 92037\\
\texttt{gperrone@westhealth.org}\\
\And Jose Unpingco\\
Gary and Mary West Health Institute\\
La Jolla, CA 92037\\
\texttt{jhunpingco@westhealth.org}\\
\And Haw-minn Lu\\
Gary and Mary West Health Institute\\
La Jolla, CA 92037\\
\texttt{hlu@westhealth.org}\\
}\maketitle
\InputIfFileExists{page_numbers.tex}{}{}
\newcommand*{\docutilsroleref}{\ref}
\newcommand*{\docutilsrolelabel}{\label}
\newcommand*\DUrolecode[1]{#1}
\providecommand*\DUrolecite[1]{\cite{#1}}
\begin{abstract}Pyvis is a Python module that enables visualizing and interactively manipulating network graphs in the Jupyter notebook, or as a standalone web application. Pyvis is built on top of the powerful and mature VisJS JavaScript library, which allows for fast and responsive interactions while also abstracting away the low-level JavaScript and HTML. This means that elements of the rendered graph visualization, such as node/edge attributes can be specified within Python and shipped to the JavaScript layer for VisJS to render. This declarative approach makes it easy to quickly explore graph visualizations and investigate data relationships. In addition, Pyvis is highly customizable so that colors, sizes, and hover tooltips can be assigned to the rendered graph. The network graph layout is controlled by a front-end physics engine that is configurable from a Python interface, allowing for the detailed placement of the graph elements. In this paper, we outline use cases for Pyvis with specific examples to highlight key features for any analysis workflow. A brief overview of Pyvis' implementation describes how the Python front-end binding uses simple Pyvis calls.\end{abstract}\keywords{networks, graphs, relationship}networks, graphs, relationship

\subsection{Introduction%
  \label{introduction}%
}

Successful Data Science is about discovering meaningful relationships in data. Visually representing these relationships using a network graph helps to accelerate understanding and make data driven decisions. Many research areas take advantage of the insight that network analysis techniques can offer. Fields in social networking, cognitive studies, telecommunications, and biological systems all leverage the applications of network theory and computation. Representing these relationships using a network graph is fundamental to all approaches, but generating an interactive and fluid graph visualization can be challenging, especially for large datasets. We introduce Pyvis, based upon the mature VisJS \DUrole{cite}{visjs} JavaScript library which enables fluid and interactive visualizations of complex network graphs. Pyvis seeks to simplify the interactive process by implementing an existing JavaScript graphics library to abstract away the low-level front end components, leaving the construction of these network data structures to Python.

The Pyvis network data structure matches the JavaScript VisJS object. This makes it easy to interpret and implement the underlying data structures from the Python layer, since the actual front end component is generated by the JavaScript library. A resulting static HTML document shows the network graph, with interactions such as dragging, zooming, hovering, and clicking. These interactions help visualize dense complex networks that are hard to explore using static graphics.

Before open-sourcing Pyvis, we used it successfully to understand relationships among hundreds of variables in a complex survey. Although we maintained an efficient data structure to represent the trends in the survey responses, we still needed a way to visualize and interact with additional metadata. Pyvis made it easy to abstract our existing data structure into nodes and edges with our desired metadata and then render the visualization with VisJS to easily identify the interrelationships. In this paper, we describe the design of Pyvis with examples showing the data structures which are rendered by VisJS.

In the following section, we demonstrate how to get up and running with Pyvis in a smaller scope by showing off the common methods of creating a network. This will also include some exposure to the customizability options that makes Pyvis so useful.

In the Layout section, we will see exactly how nodes and edges can be spatially specified by interacting with various physics parameters interpreted by the front end engine.

Integrations with Jupyter and NetworkX will be presented to establish Pyvis compatibility with popular data science workflows.

Finally, a thought out example will include the interpretation of a practical Game of Thrones relationship dataset to demonstrate a Pyvis use case from the ground up. This minimal example will be a base case for the features that Pyvis supports.

\subsection{Pyvis Usage%
  \label{pyvis-usage}%
}

Installing Pyvis is straight-forward with details at the project documentation website \DUrole{cite}{pyvis}. All of the following examples will utilize familiar Python data structures with some connections to the popular and powerful NetworkX package \DUrole{cite}{networkx}. The basic \texttt{Network} class is the container for graph and front end properties. All networks must be instantiated as a \texttt{Network} class instance:\vspace{1mm}
\begin{Verbatim}[commandchars=\\\{\},fontsize=\footnotesize]
\PY{k+kn}{from} \PY{n+nn}{pyvis}\PY{n+nn}{.}\PY{n+nn}{network} \PY{k+kn}{import} \PY{n}{Network}
\PY{n}{g} \PY{o}{=} \PY{n}{Network}\PY{p}{(}\PY{p}{)}
\end{Verbatim}
\vspace{1mm}
Nodes can be added by providing an integer or string \texttt{id} and an optional label.\vspace{1mm}
\begin{Verbatim}[commandchars=\\\{\},fontsize=\footnotesize]
\PY{n}{g}\PY{o}{.}\PY{n}{add\PYZus{}node}\PY{p}{(}\PY{l+m+mi}{1}\PY{p}{)}
\PY{n}{g}\PY{o}{.}\PY{n}{add\PYZus{}node}\PY{p}{(}\PY{l+m+mi}{2}\PY{p}{)}
\PY{n+nb}{print}\PY{p}{(}\PY{n}{g}\PY{p}{)}
\end{Verbatim}
\vspace{1mm}
\vspace{1mm}
\begin{Verbatim}[commandchars=\\\{\},fontsize=\footnotesize]
\PY{p}{\PYZob{}}
 \PY{n+nt}{\PYZdq{}Nodes\PYZdq{}}\PY{p}{:} \PY{p}{[}
     \PY{l+m+mi}{1}\PY{p}{,}
     \PY{l+m+mi}{2}
 \PY{p}{]}\PY{p}{,}
 \PY{n+nt}{\PYZdq{}Edges\PYZdq{}}\PY{p}{:} \PY{p}{[}\PY{p}{]}\PY{p}{,}
 \PY{n+nt}{\PYZdq{}Height\PYZdq{}}\PY{p}{:} \PY{l+s+s2}{\PYZdq{}500px\PYZdq{}}\PY{p}{,}
 \PY{n+nt}{\PYZdq{}Width\PYZdq{}}\PY{p}{:} \PY{l+s+s2}{\PYZdq{}500px\PYZdq{}}
\PY{p}{\PYZcb{}}
\end{Verbatim}
\vspace{1mm}
The \texttt{add\_nodes} method consumes a list of nodes:\vspace{1mm}
\begin{Verbatim}[commandchars=\\\{\},fontsize=\footnotesize]
\PY{n}{nodes} \PY{o}{=} \PY{p}{[}\PY{l+s+s2}{\PYZdq{}}\PY{l+s+s2}{a}\PY{l+s+s2}{\PYZdq{}}\PY{p}{,} \PY{l+s+s2}{\PYZdq{}}\PY{l+s+s2}{b}\PY{l+s+s2}{\PYZdq{}}\PY{p}{,} \PY{l+s+s2}{\PYZdq{}}\PY{l+s+s2}{c}\PY{l+s+s2}{\PYZdq{}}\PY{p}{,} \PY{l+s+s2}{\PYZdq{}}\PY{l+s+s2}{d}\PY{l+s+s2}{\PYZdq{}}\PY{p}{]}
\PY{n}{g}\PY{o}{.}\PY{n}{add\PYZus{}nodes}\PY{p}{(}\PY{n}{nodes}\PY{p}{)}
\PY{n}{g}\PY{o}{.}\PY{n}{add\PYZus{}nodes}\PY{p}{(}\PY{l+s+s2}{\PYZdq{}}\PY{l+s+s2}{hello}\PY{l+s+s2}{\PYZdq{}}\PY{p}{)}
\end{Verbatim}
\vspace{1mm}
Keyword arguments can be used to add properties to the nodes in \texttt{Network}:\vspace{1mm}
\begin{Verbatim}[commandchars=\\\{\},fontsize=\footnotesize]
\PY{n}{g} \PY{o}{=} \PY{n}{Network}\PY{p}{(}\PY{p}{)}
\PY{n}{g}\PY{o}{.}\PY{n}{add\PYZus{}nodes}\PY{p}{(}
   \PY{p}{[}\PY{l+m+mi}{1}\PY{p}{,}\PY{l+m+mi}{2}\PY{p}{,}\PY{l+m+mi}{3}\PY{p}{]}\PY{p}{,}
   \PY{n}{value}\PY{o}{=}\PY{p}{[}\PY{l+m+mi}{10}\PY{p}{,} \PY{l+m+mi}{100}\PY{p}{,} \PY{l+m+mi}{400}\PY{p}{]}\PY{p}{,} \PY{c+c1}{\PYZsh{} values adjust node size}
   \PY{n}{x}\PY{o}{=}\PY{p}{[}\PY{l+m+mf}{21.4}\PY{p}{,} \PY{l+m+mf}{154.2}\PY{p}{,} \PY{l+m+mf}{11.2}\PY{p}{]}\PY{p}{,}
   \PY{n}{y}\PY{o}{=}\PY{p}{[}\PY{l+m+mf}{100.2}\PY{p}{,} \PY{l+m+mf}{23.54}\PY{p}{,} \PY{l+m+mf}{32.1}\PY{p}{]}\PY{p}{,}
   \PY{n}{label}\PY{o}{=}\PY{p}{[}\PY{l+s+s2}{\PYZdq{}}\PY{l+s+s2}{NODE 1}\PY{l+s+s2}{\PYZdq{}}\PY{p}{,} \PY{l+s+s2}{\PYZdq{}}\PY{l+s+s2}{NODE 2}\PY{l+s+s2}{\PYZdq{}}\PY{p}{,} \PY{l+s+s2}{\PYZdq{}}\PY{l+s+s2}{NODE 3}\PY{l+s+s2}{\PYZdq{}}\PY{p}{]}\PY{p}{,}
   \PY{n}{color}\PY{o}{=}\PY{p}{[}\PY{l+s+s2}{\PYZdq{}}\PY{l+s+s2}{\PYZsh{}00ff1e}\PY{l+s+s2}{\PYZdq{}}\PY{p}{,} \PY{l+s+s2}{\PYZdq{}}\PY{l+s+s2}{\PYZsh{}162347}\PY{l+s+s2}{\PYZdq{}}\PY{p}{,} \PY{l+s+s2}{\PYZdq{}}\PY{l+s+s2}{\PYZsh{}dd4b39}\PY{l+s+s2}{\PYZdq{}}\PY{p}{]}
\PY{p}{)}
\PY{n}{g}\PY{o}{.}\PY{n}{show}\PY{p}{(}\PY{l+s+s2}{\PYZdq{}}\PY{l+s+s2}{example.html}\PY{l+s+s2}{\PYZdq{}}\PY{p}{)}
\end{Verbatim}
\vspace{1mm}
\noindent\makebox[\columnwidth][l]{\includegraphics[]{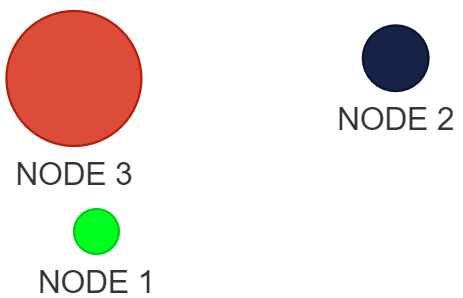}}
\begin{DUlineblock}{0em}
\item[] The following node properties influence the resulting visualization:
\end{DUlineblock}

\begin{itemize}
\item size - The raw circumference of a single node
\item 

value - Circumference of node but scaled according to all values
\item 

title - The title displays over each node while mousing over it
\item 

x - X coordinate of node for custom layouts
\item 

y - Y coordinate of node for custom layouts
\item 

label - A label appearing under each node
\item 

color - The color of the node\end{itemize}

\begin{DUlineblock}{0em}
\item[] Nodes must exist in the network instance in order to add edges
\end{DUlineblock}
\vspace{1mm}
\begin{Verbatim}[commandchars=\\\{\},fontsize=\footnotesize]
\PY{n}{g}\PY{o}{.}\PY{n}{add\PYZus{}edge}\PY{p}{(}\PY{l+m+mi}{1}\PY{p}{,} \PY{l+m+mi}{2}\PY{p}{)}
\PY{c+c1}{\PYZsh{} will adjust edge thickness}
\PY{n}{g}\PY{o}{.}\PY{n}{add\PYZus{}edge}\PY{p}{(}\PY{l+m+mi}{2}\PY{p}{,} \PY{l+m+mi}{3}\PY{p}{,} \PY{n}{weight}\PY{o}{=}\PY{l+m+mi}{5}\PY{p}{)}
\end{Verbatim}
\vspace{1mm}
Edges can be added all at once by supplying a list of tuples to a call to \DUroletitlereference{add\_edges()}. The following is an equivalent result:\vspace{1mm}
\begin{Verbatim}[commandchars=\\\{\},fontsize=\footnotesize]
\PY{n}{g}\PY{o}{.}\PY{n}{add\PYZus{}edges}\PY{p}{(}\PY{p}{[}\PY{p}{(}\PY{l+m+mi}{1}\PY{p}{,} \PY{l+m+mi}{2}\PY{p}{)}\PY{p}{,} \PY{p}{(}\PY{l+m+mi}{2}\PY{p}{,} \PY{l+m+mi}{3}\PY{p}{,} \PY{l+m+mi}{5}\PY{p}{)}\PY{p}{]}\PY{p}{)}
\PY{n}{g}\PY{o}{.}\PY{n}{show}\PY{p}{(}\PY{l+s+s2}{\PYZdq{}}\PY{l+s+s2}{example.html}\PY{l+s+s2}{\PYZdq{}}\PY{p}{)}
\end{Verbatim}
\vspace{1mm}
\noindent\makebox[\columnwidth][l]{\includegraphics[]{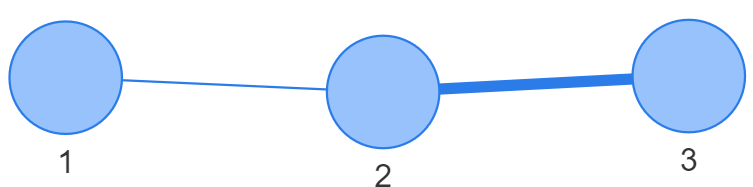}}
\begin{DUlineblock}{0em}
\item[] Notice how an optional element is included in the 3-tuple above (2, 3, 5) representing the weight of the edge. This additional edge data allows for expressing weighted networks and is clearly noticeable in the visualization.
\end{DUlineblock}

\subsection{Layout%
  \label{layout}%
}

\begin{DUlineblock}{0em}
\item[] In situations where your network involves complex connections, Pyvis allows you to manually explore these relationships with intuitive mouse interactions. Nodes can be dragged into more visible positions if the view is obstructed.
\item[] All of this is made possible by the front end engine provided by VisJS. Their extensive documentation defines several options for supplying layout and physics configurations to instances of a network. These physics options are fundamental to VisJS, so tweaking the physics of the rendered simulation is as simple as providing the parameters to the specific solver.
\end{DUlineblock}
The physics options dictates how a user can interact with the objects in the graph. The intent of the physic options is to make manipulating graph objects feel more intuitive when moving nodes around. As an example, the user can manipulate a portion of a graph that is densely populated to view a graph segment of the interest more clearly. VisJS implements several physical simulations such as Barnes Hut \DUrole{cite}{barneshut}. Others are mentioned in the VisJS documentation \DUrole{cite}{visjsphysics}.
\begin{DUlineblock}{0em}
\item[] We can configure the physics engine from within Pyvis:
\end{DUlineblock}
\vspace{1mm}
\begin{Verbatim}[commandchars=\\\{\},fontsize=\footnotesize]
\PY{n}{g} \PY{o}{=} \PY{n}{Network}\PY{p}{(}\PY{p}{)}
\PY{c+c1}{\PYZsh{} physics solvers supported:}
\PY{c+c1}{\PYZsh{} barnesHut, forceAtlas2Based, repulsion,}
\PY{c+c1}{\PYZsh{} hierarchicalRepulsion}
\PY{n}{g}\PY{o}{.}\PY{n}{barnes\PYZus{}hut}\PY{p}{(}
 \PY{n}{gravity}\PY{o}{=}\PY{o}{\PYZhy{}}\PY{l+m+mi}{80000}\PY{p}{,}
 \PY{n}{central\PYZus{}gravity}\PY{o}{=}\PY{l+m+mf}{0.3}\PY{p}{,}
 \PY{n}{spring\PYZus{}length}\PY{o}{=}\PY{l+m+mi}{250}\PY{p}{,}
 \PY{n}{spring\PYZus{}strength}\PY{o}{=}\PY{l+m+mf}{0.001}\PY{p}{,}
 \PY{n}{damping}\PY{o}{=}\PY{l+m+mf}{0.09}\PY{p}{,}
 \PY{n}{overlap}\PY{o}{=}\PY{l+m+mi}{0}\PY{p}{,}
\PY{p}{)}
\PY{n+nb}{print}\PY{p}{(}\PY{n}{g}\PY{o}{.}\PY{n}{options}\PY{o}{.}\PY{n}{physics}\PY{p}{)}
\PY{p}{\PYZob{}}\PY{l+s+s1}{\PYZsq{}}\PY{l+s+s1}{enabled}\PY{l+s+s1}{\PYZsq{}}\PY{p}{:} \PY{k+kc}{True}\PY{p}{,}
\PY{l+s+s1}{\PYZsq{}}\PY{l+s+s1}{stabilization}\PY{l+s+s1}{\PYZsq{}}\PY{p}{:}
\PY{o}{\PYZlt{}}\PY{n}{pyvis}\PY{o}{.}\PY{n}{physics}\PY{o}{.}\PY{n}{Physics}\PY{o}{.}\PY{n}{Stabilization}
\PY{n+nb}{object} \PY{n}{at} \PY{l+m+mh}{0x7f99e6a03f90}\PY{o}{\PYZgt{}}\PY{p}{,}
\PY{l+s+s1}{\PYZsq{}}\PY{l+s+s1}{barnesHut}\PY{l+s+s1}{\PYZsq{}}\PY{p}{:} \PY{o}{\PYZlt{}}\PY{n}{pyvis}\PY{o}{.}\PY{n}{physics}\PY{o}{.}\PY{n}{Physics}\PY{o}{.}\PY{n}{barnesHut}
\PY{n+nb}{object} \PY{n}{at} \PY{l+m+mh}{0x7f99e6de3710}\PY{o}{\PYZgt{}}\PY{p}{\PYZcb{}}
\end{Verbatim}
\vspace{1mm}

\begin{DUlineblock}{0em}
\item[] In order to avoid the scenario of \textquotedbl{}guessing\textquotedbl{} parameter values for an optimal network physics configuration, VisJS offers a useful interaction for experimenting with theses values.
\item[] These interactions are enabled via Pyvis:
\end{DUlineblock}
\vspace{1mm}
\begin{Verbatim}[commandchars=\\\{\},fontsize=\footnotesize]
\PY{c+c1}{\PYZsh{} choose to only show the physics options}
\PY{n}{g}\PY{o}{.}\PY{n}{show\PYZus{}buttons}\PY{p}{(}\PY{n}{filter\PYZus{}}\PY{o}{=}\PY{p}{[}\PY{l+s+s2}{\PYZdq{}}\PY{l+s+s2}{physics}\PY{l+s+s2}{\PYZdq{}}\PY{p}{]}\PY{p}{)}
\end{Verbatim}
\vspace{1mm}
\noindent\makebox[\columnwidth][l]{\includegraphics[width=\columnwidth]{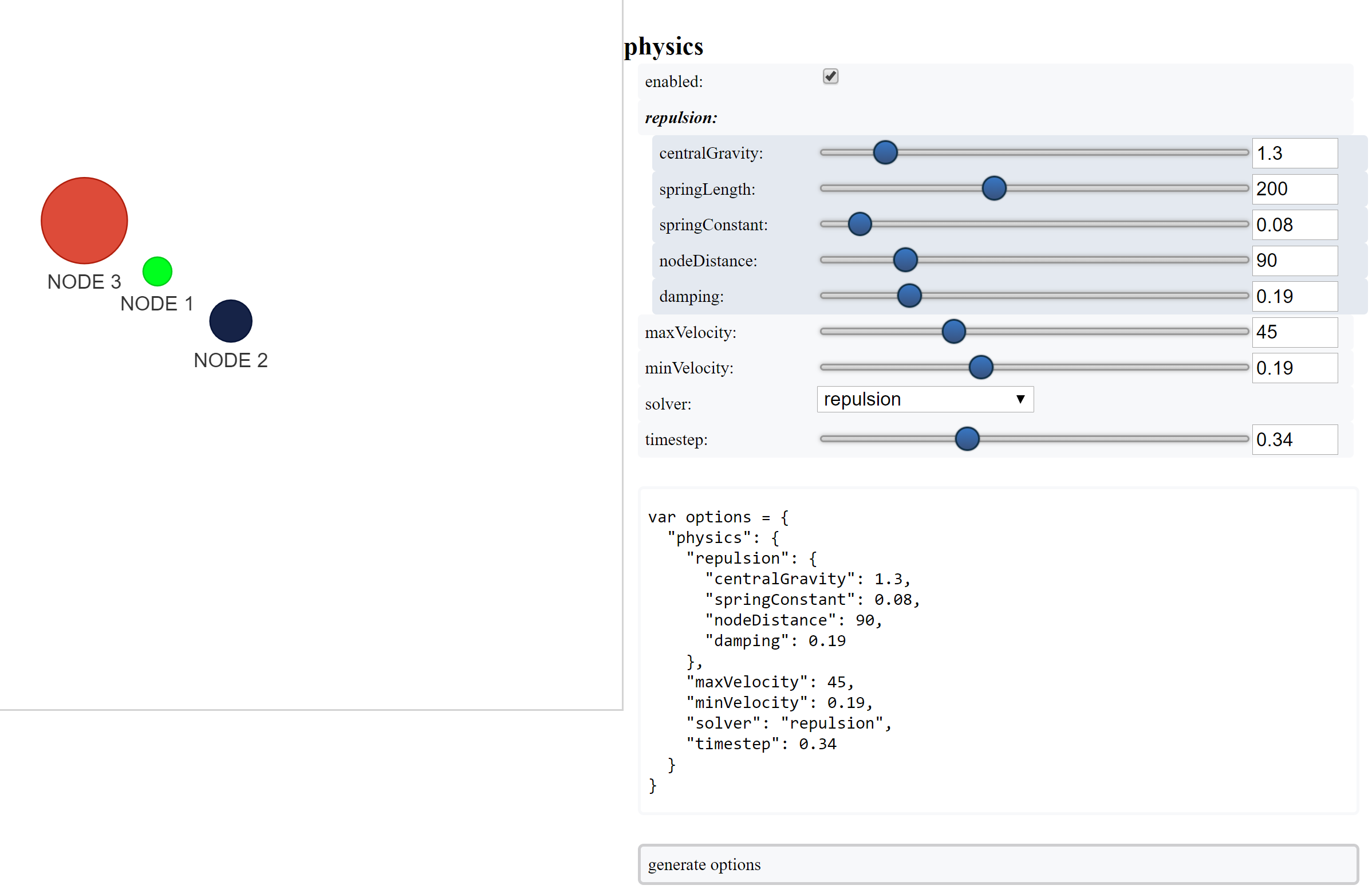}}
\begin{DUlineblock}{0em}
\item[] Here, we choose to display the options for the physics component of the network. Omitting a filter in the call will display the configuration of the entire network including nodes, edges, layout, and interaction. The JSON options displayed in the visualization represent the current configuration depending on the displayed sliders. You can copy/paste those options to supply your network with custom settings:
\end{DUlineblock}
\vspace{1mm}
\begin{Verbatim}[commandchars=\\\{\},fontsize=\footnotesize]
\PY{n}{g}\PY{o}{.}\PY{n}{set\PYZus{}options}\PY{p}{(}
   \PY{l+s+sd}{\PYZdq{}\PYZdq{}\PYZdq{}}
\PY{l+s+sd}{   var options = \PYZob{}}
\PY{l+s+sd}{      \PYZdq{}physics\PYZdq{}: \PYZob{}}
\PY{l+s+sd}{         \PYZdq{}repulsion\PYZdq{}: \PYZob{}}
\PY{l+s+sd}{            \PYZdq{}centralGravity\PYZdq{}: 1.3,}
\PY{l+s+sd}{            \PYZdq{}springConstant\PYZdq{}: 0.08,}
\PY{l+s+sd}{            \PYZdq{}nodeDistance\PYZdq{}: 90,}
\PY{l+s+sd}{            \PYZdq{}damping\PYZdq{}: 0.19}
\PY{l+s+sd}{         \PYZcb{},}
\PY{l+s+sd}{         \PYZdq{}maxVelocity\PYZdq{}: 45,}
\PY{l+s+sd}{         \PYZdq{}minVelocity\PYZdq{}: 0.19,}
\PY{l+s+sd}{         \PYZdq{}solver\PYZdq{}: \PYZdq{}repulsion\PYZdq{},}
\PY{l+s+sd}{         \PYZdq{}timestep\PYZdq{}: 0.34}
\PY{l+s+sd}{      \PYZcb{}}
\PY{l+s+sd}{   \PYZcb{}}
\PY{l+s+sd}{   \PYZdq{}\PYZdq{}\PYZdq{}}
\PY{p}{)}
\PY{n+nb}{print}\PY{p}{(}\PY{n}{g}\PY{o}{.}\PY{n}{options}\PY{p}{)}
\end{Verbatim}
\vspace{1mm}
\vspace{1mm}
\begin{Verbatim}[commandchars=\\\{\},fontsize=\footnotesize]
\PY{p}{\PYZob{}}\PY{l+s+s1}{\PYZsq{}}\PY{l+s+s1}{physics}\PY{l+s+s1}{\PYZsq{}}\PY{p}{:} \PY{p}{\PYZob{}}\PY{l+s+s1}{\PYZsq{}}\PY{l+s+s1}{repulsion}\PY{l+s+s1}{\PYZsq{}}\PY{p}{:} \PY{p}{\PYZob{}}\PY{l+s+s1}{\PYZsq{}}\PY{l+s+s1}{centralGravity}\PY{l+s+s1}{\PYZsq{}}\PY{p}{:} \PY{l+m+mf}{1.3}\PY{p}{,}
\PY{l+s+s1}{\PYZsq{}}\PY{l+s+s1}{springConstant}\PY{l+s+s1}{\PYZsq{}}\PY{p}{:} \PY{l+m+mf}{0.08}\PY{p}{,}
\PY{l+s+s1}{\PYZsq{}}\PY{l+s+s1}{nodeDistance}\PY{l+s+s1}{\PYZsq{}}\PY{p}{:} \PY{l+m+mi}{90}\PY{p}{,}
\PY{l+s+s1}{\PYZsq{}}\PY{l+s+s1}{damping}\PY{l+s+s1}{\PYZsq{}}\PY{p}{:} \PY{l+m+mf}{0.19}\PY{p}{\PYZcb{}}\PY{p}{,}
\PY{l+s+s1}{\PYZsq{}}\PY{l+s+s1}{maxVelocity}\PY{l+s+s1}{\PYZsq{}}\PY{p}{:} \PY{l+m+mi}{45}\PY{p}{,}
\PY{l+s+s1}{\PYZsq{}}\PY{l+s+s1}{minVelocity}\PY{l+s+s1}{\PYZsq{}}\PY{p}{:} \PY{l+m+mf}{0.19}\PY{p}{,}
\PY{l+s+s1}{\PYZsq{}}\PY{l+s+s1}{solver}\PY{l+s+s1}{\PYZsq{}}\PY{p}{:} \PY{l+s+s1}{\PYZsq{}}\PY{l+s+s1}{repulsion}\PY{l+s+s1}{\PYZsq{}}\PY{p}{,}
\PY{l+s+s1}{\PYZsq{}}\PY{l+s+s1}{timestep}\PY{l+s+s1}{\PYZsq{}}\PY{p}{:} \PY{l+m+mf}{0.34}\PY{p}{\PYZcb{}}\PY{p}{\PYZcb{}}
\end{Verbatim}
\vspace{1mm}

\begin{DUlineblock}{0em}
\item[] The methods of a \texttt{Network} instance construct an internal structure compatible with VisJS, demonstrated by the consistent pattern of JSON outputs seen above.
\end{DUlineblock}

\subsection{NetworkX Support%
  \label{networkx-support}%
}
Although Pyvis supports its own methods for constructing a network data structure, you might feel more comfortable using the more established and dedicated NetworkX package. Pyvis allows you to define a NetworkX graph instance to then supply it to Pyvis.\vspace{1mm}
\begin{Verbatim}[commandchars=\\\{\},fontsize=\footnotesize]
\PY{k+kn}{import} \PY{n+nn}{networkx} \PY{k}{as} \PY{n+nn}{nx}
\PY{k+kn}{from} \PY{n+nn}{pyvis}\PY{n+nn}{.}\PY{n+nn}{network} \PY{k+kn}{import} \PY{n}{Network}

\PY{n}{nxg} \PY{o}{=} \PY{n}{nx}\PY{o}{.}\PY{n}{random\PYZus{}tree}\PY{p}{(}\PY{l+m+mi}{20}\PY{p}{)}
\PY{n}{g}\PY{o}{=}\PY{n}{Network}\PY{p}{(}\PY{n}{directed}\PY{o}{=}\PY{k+kc}{True}\PY{p}{)}
\PY{n}{g}\PY{o}{.}\PY{n}{from\PYZus{}nx}\PY{p}{(}\PY{n}{nxg}\PY{p}{)}
\PY{n}{g}\PY{o}{.}\PY{n}{show}\PY{p}{(}\PY{l+s+s2}{\PYZdq{}}\PY{l+s+s2}{networkx.html}\PY{l+s+s2}{\PYZdq{}}\PY{p}{)}
\end{Verbatim}
\vspace{1mm}
\noindent\makebox[\columnwidth][l]{\includegraphics[width=\columnwidth]{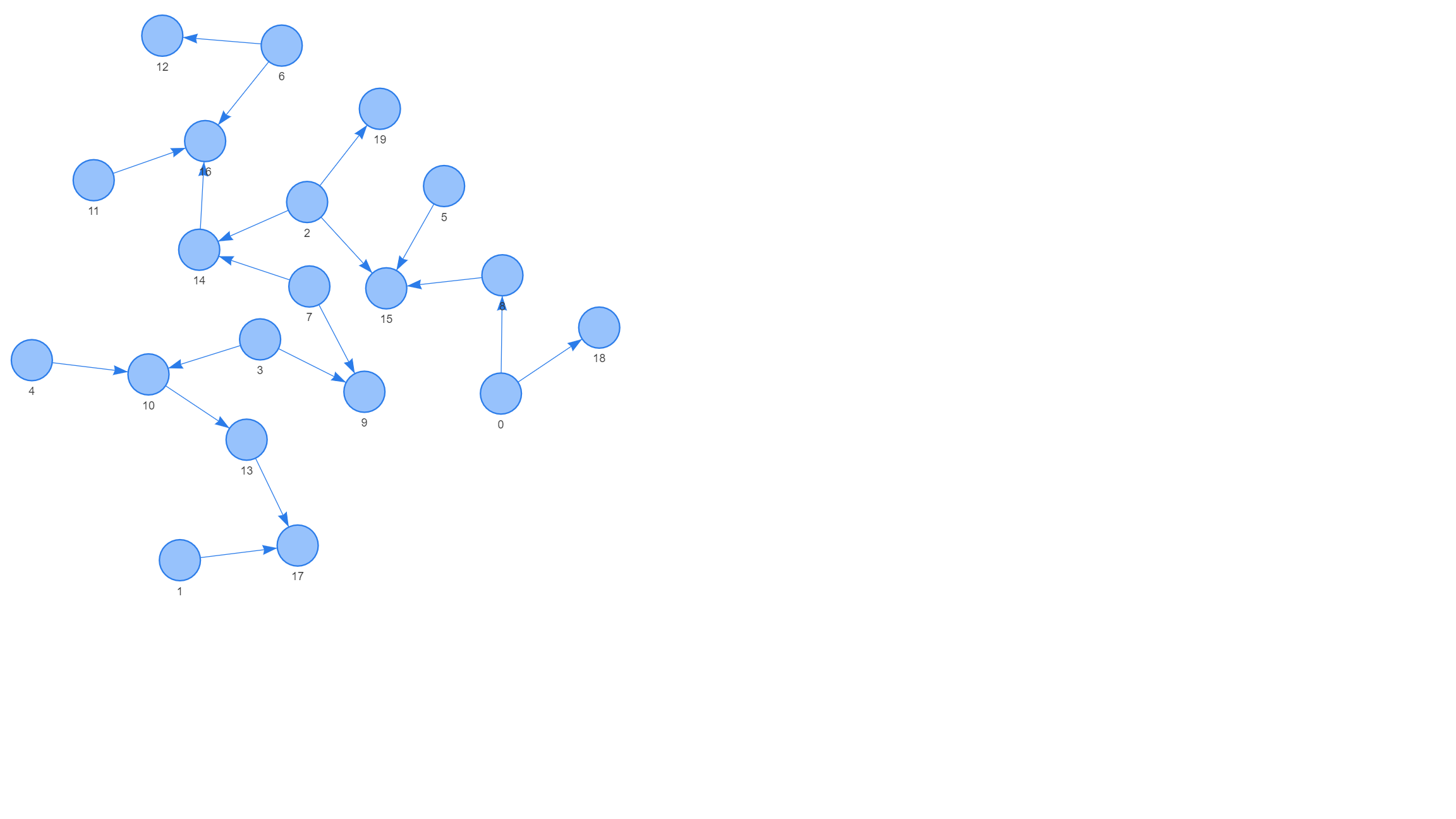}}
\begin{DUlineblock}{0em}
\item[] Pyvis current behavior recognizes the basic topology of a NetworkX graph, not accounting for any custom attributes provided. Any other attributes like node color, size, and layout would need to be manually added to the resulting Pyvis graph.
\item[] Future plans are to fully integrate NetworkX graphs to fully interpret them, preserving attributes in the resulting Pyvis visualizations.
\end{DUlineblock}

\subsection{Jupyter Support%
  \label{jupyter-support}%
}
For efficient prototyping of visualized graphs, Pyvis aims to utilize Jupyter's front-end IFrame features to embed the graph in a notebook output cell.
With that in mind, embedding a Pyvis visualization into a Jupyter notebook is essentially the same as described above. The only difference is that one should pass in a notebook argument during instantiation. The result of the visualization is shown in the output cell below the \texttt{show()} invocation. Pyvis upon the call to \texttt{show()} writes the HTML that serves an IFrame, which displays the result in the output cell.
|\noindent\makebox[\columnwidth][l]{\includegraphics[width=\columnwidth]{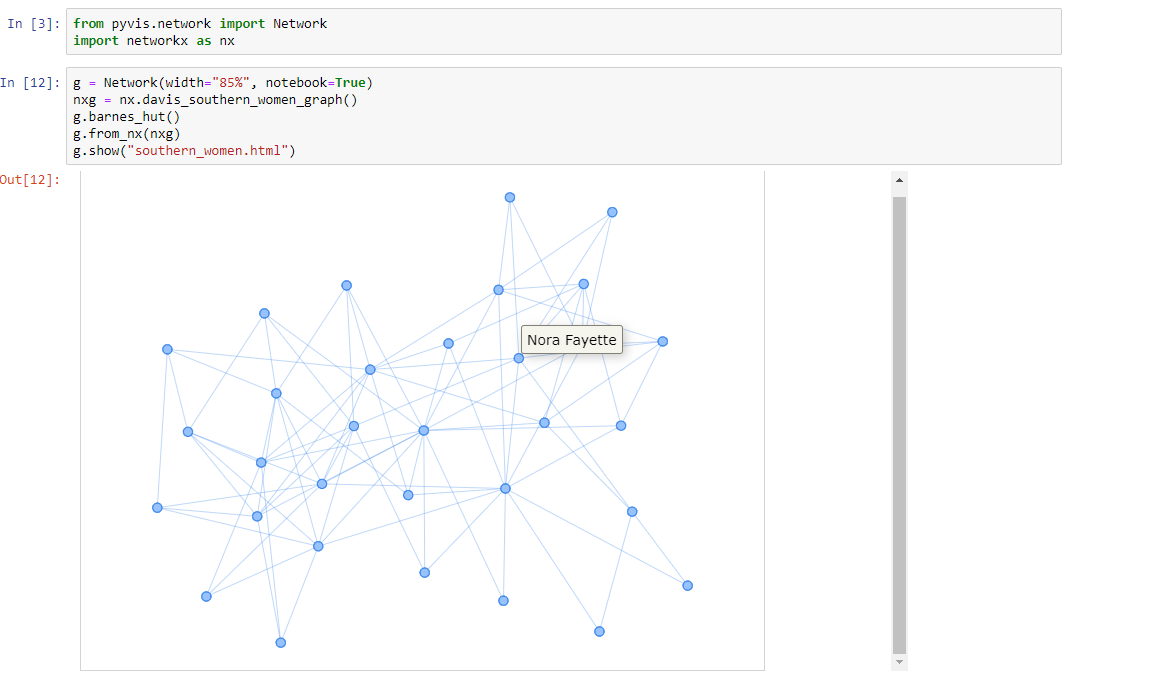}}
\begin{DUlineblock}{0em}
\item[] One thing to keep in mind is that an HTML file is always generated due to the dependence on the VisJS JavaScript bindings.
\end{DUlineblock}

\subsection{Example%
  \label{example}%
}

\begin{DUlineblock}{0em}
\item[] To get a better understanding of the flow of a typical Pyvis network visualization, we can take a look at the following code snippet to show off a typical application of the features. I have taken a Game of Thrones dataset (\DUrole{cite}{gthrones} Storm of Swords Dataset) defining the relationships between characters and the frequencies between them to create a network to naturally express this. Specifically, it is a csv file containing pairs of characters and a weight between them.
\end{DUlineblock}
\vspace{1mm}
\begin{Verbatim}[commandchars=\\\{\},fontsize=\footnotesize]
\PY{k+kn}{from} \PY{n+nn}{pyvis}\PY{n+nn}{.}\PY{n+nn}{network} \PY{k+kn}{import} \PY{n}{Network}
\PY{k+kn}{import} \PY{n+nn}{pandas} \PY{k}{as} \PY{n+nn}{pd}

\PY{n}{got\PYZus{}net} \PY{o}{=} \PY{n}{Network}\PY{p}{(}
   \PY{n}{height}\PY{o}{=}\PY{l+s+s2}{\PYZdq{}}\PY{l+s+s2}{750px}\PY{l+s+s2}{\PYZdq{}}\PY{p}{,}
   \PY{n}{width}\PY{o}{=}\PY{l+s+s2}{\PYZdq{}}\PY{l+s+s2}{100}\PY{l+s+s2}{\PYZpc{}}\PY{l+s+s2}{\PYZdq{}}\PY{p}{,}
   \PY{n}{bgcolor}\PY{o}{=}\PY{l+s+s2}{\PYZdq{}}\PY{l+s+s2}{\PYZsh{}222222}\PY{l+s+s2}{\PYZdq{}}\PY{p}{,}
   \PY{n}{font\PYZus{}color}\PY{o}{=}\PY{l+s+s2}{\PYZdq{}}\PY{l+s+s2}{white}\PY{l+s+s2}{\PYZdq{}}
\PY{p}{)}

\PY{c+c1}{\PYZsh{} set the physics layout of the network}
\PY{n}{got\PYZus{}net}\PY{o}{.}\PY{n}{barnes\PYZus{}hut}\PY{p}{(}\PY{p}{)}
\PY{n}{got\PYZus{}data} \PY{o}{=} \PY{n}{pd}\PY{o}{.}\PY{n}{read\PYZus{}csv}\PY{p}{(}\PY{l+s+s2}{\PYZdq{}}\PY{l+s+s2}{stormofswords.csv}\PY{l+s+s2}{\PYZdq{}}\PY{p}{)}

\PY{n}{sources} \PY{o}{=} \PY{n}{got\PYZus{}data}\PY{p}{[}\PY{l+s+s1}{\PYZsq{}}\PY{l+s+s1}{Source}\PY{l+s+s1}{\PYZsq{}}\PY{p}{]}
\PY{n}{targets} \PY{o}{=} \PY{n}{got\PYZus{}data}\PY{p}{[}\PY{l+s+s1}{\PYZsq{}}\PY{l+s+s1}{Target}\PY{l+s+s1}{\PYZsq{}}\PY{p}{]}
\PY{n}{weights} \PY{o}{=} \PY{n}{got\PYZus{}data}\PY{p}{[}\PY{l+s+s1}{\PYZsq{}}\PY{l+s+s1}{Weight}\PY{l+s+s1}{\PYZsq{}}\PY{p}{]}

\PY{n}{edge\PYZus{}data} \PY{o}{=} \PY{n+nb}{zip}\PY{p}{(}\PY{n}{sources}\PY{p}{,} \PY{n}{targets}\PY{p}{,} \PY{n}{weights}\PY{p}{)}

\PY{k}{for} \PY{n}{e} \PY{o+ow}{in} \PY{n}{edge\PYZus{}data}\PY{p}{:}
   \PY{n}{src} \PY{o}{=} \PY{n}{e}\PY{p}{[}\PY{l+m+mi}{0}\PY{p}{]}
   \PY{n}{dst} \PY{o}{=} \PY{n}{e}\PY{p}{[}\PY{l+m+mi}{1}\PY{p}{]}
   \PY{n}{w} \PY{o}{=} \PY{n}{e}\PY{p}{[}\PY{l+m+mi}{2}\PY{p}{]}

   \PY{n}{got\PYZus{}net}\PY{o}{.}\PY{n}{add\PYZus{}node}\PY{p}{(}\PY{n}{src}\PY{p}{,} \PY{n}{src}\PY{p}{,} \PY{n}{title}\PY{o}{=}\PY{n}{src}\PY{p}{)}
   \PY{n}{got\PYZus{}net}\PY{o}{.}\PY{n}{add\PYZus{}node}\PY{p}{(}\PY{n}{dst}\PY{p}{,} \PY{n}{dst}\PY{p}{,} \PY{n}{title}\PY{o}{=}\PY{n}{dst}\PY{p}{)}
   \PY{n}{got\PYZus{}net}\PY{o}{.}\PY{n}{add\PYZus{}edge}\PY{p}{(}\PY{n}{src}\PY{p}{,} \PY{n}{dst}\PY{p}{,} \PY{n}{value}\PY{o}{=}\PY{n}{w}\PY{p}{)}

\PY{n}{neighbor\PYZus{}map} \PY{o}{=} \PY{n}{got\PYZus{}net}\PY{o}{.}\PY{n}{get\PYZus{}adj\PYZus{}list}\PY{p}{(}\PY{p}{)}

\PY{c+c1}{\PYZsh{} add neighbor data to node hover data}
\PY{k}{for} \PY{n}{node} \PY{o+ow}{in} \PY{n}{got\PYZus{}net}\PY{o}{.}\PY{n}{nodes}\PY{p}{:}
   \PY{n}{node}\PY{p}{[}\PY{l+s+s2}{\PYZdq{}}\PY{l+s+s2}{title}\PY{l+s+s2}{\PYZdq{}}\PY{p}{]} \PY{o}{+}\PY{o}{=} \PY{l+s+s2}{\PYZdq{}}\PY{l+s+s2}{ Neighbors:\PYZlt{}br\PYZgt{}}\PY{l+s+s2}{\PYZdq{}} \PY{o}{+} \PYZbs{}
           \PY{l+s+s2}{\PYZdq{}}\PY{l+s+s2}{\PYZlt{}br\PYZgt{}}\PY{l+s+s2}{\PYZdq{}}\PY{o}{.}\PY{n}{join}\PY{p}{(}\PY{n}{neighbor\PYZus{}map}\PY{p}{[}\PY{n}{node}\PY{p}{[}\PY{l+s+s2}{\PYZdq{}}\PY{l+s+s2}{id}\PY{l+s+s2}{\PYZdq{}}\PY{p}{]}\PY{p}{]}\PY{p}{)}
   \PY{n}{node}\PY{p}{[}\PY{l+s+s2}{\PYZdq{}}\PY{l+s+s2}{value}\PY{l+s+s2}{\PYZdq{}}\PY{p}{]} \PY{o}{=} \PY{n+nb}{len}\PY{p}{(}\PY{n}{neighbor\PYZus{}map}\PY{p}{[}\PY{n}{node}\PY{p}{[}\PY{l+s+s2}{\PYZdq{}}\PY{l+s+s2}{id}\PY{l+s+s2}{\PYZdq{}}\PY{p}{]}\PY{p}{]}\PY{p}{)}

\PY{n}{got\PYZus{}net}\PY{o}{.}\PY{n}{show}\PY{p}{(}\PY{l+s+s2}{\PYZdq{}}\PY{l+s+s2}{gameofthrones.html}\PY{l+s+s2}{\PYZdq{}}\PY{p}{)}
\end{Verbatim}
\vspace{1mm}
\noindent\makebox[\columnwidth][l]{\includegraphics[width=\columnwidth]{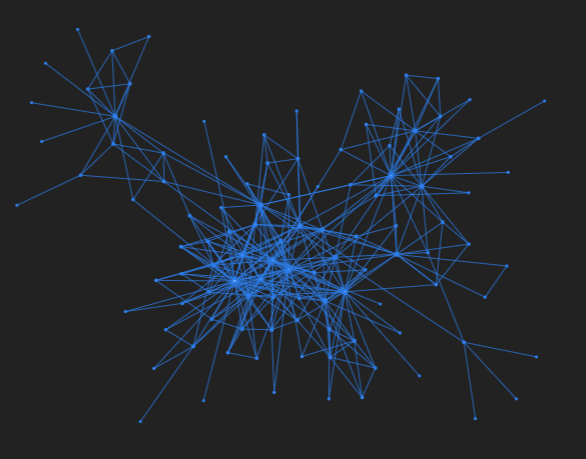}}
\begin{DUlineblock}{0em}
\item[] At a glance, the resulting relationship network looks too intertwined to make any practical conclusions. However, the beauty of Pyvis is that each and every component of the network can be focused. For example, zooming in to a dense portion of the network, we can hover over a particular node to get a glimpse of the scenario:
\end{DUlineblock}
\noindent\makebox[\columnwidth][l]{\includegraphics[width=\columnwidth]{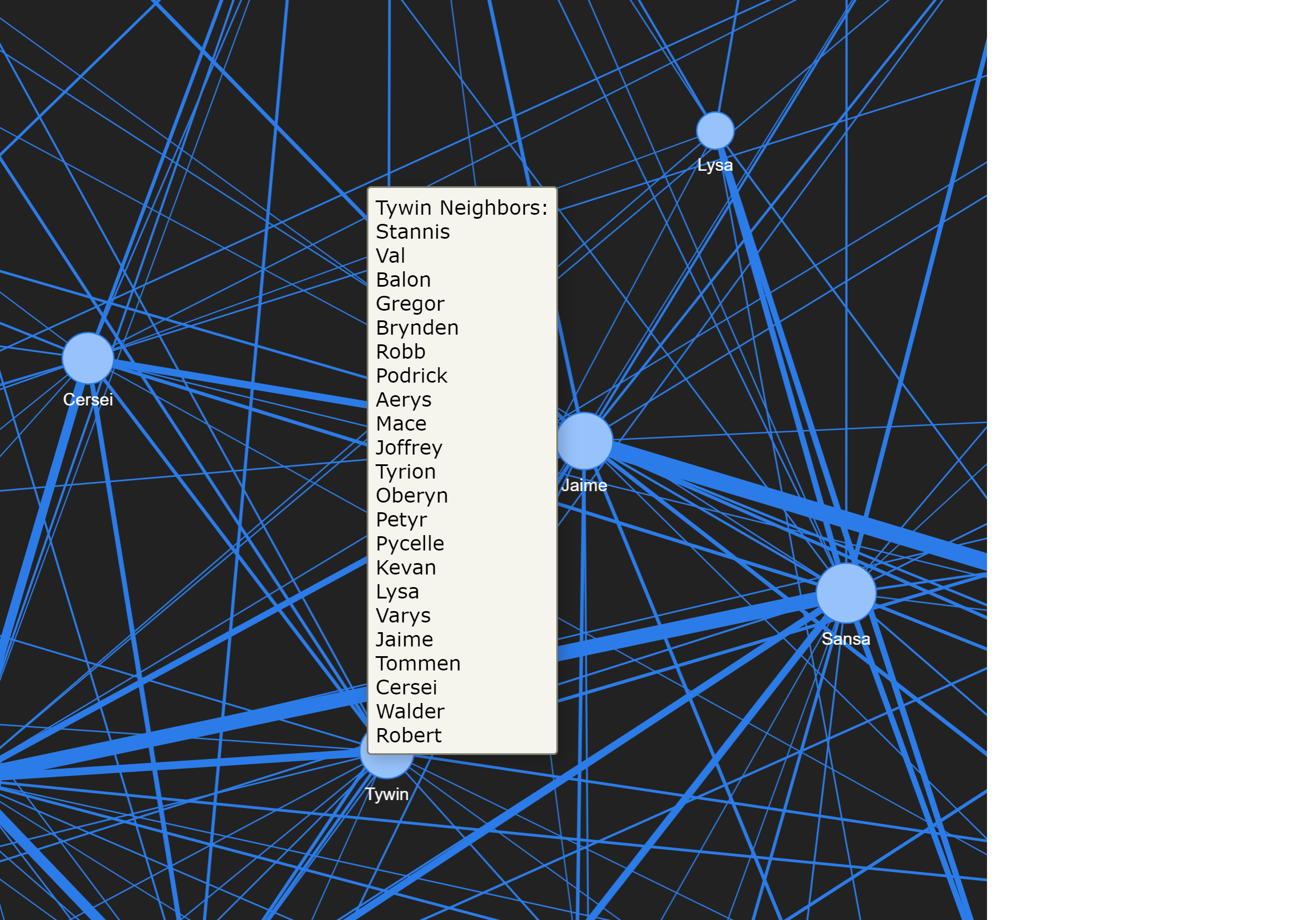}}
\begin{DUlineblock}{0em}
\item[] This hover tooltip offers the context behind a particular node. We can see the immediate neighbors for each and every node since we provided a \DUroletitlereference{title} attribute during the network construction. This simple example can be expanded upon to create more custom interactions tailored to specific needs of a dataset.
\item[] The network also uses weights. By providing a \DUroletitlereference{value} attribute to each node we can see these values being represented by a node's size. In the code I used the amount of neighbors to dictate the node weight. This is a strong visual cue which makes it easy to see which nodes have the most connections.
\item[] The edge weights are assigned in a similar manner, although the dataset already provided the connection strength between nodes. These edge weights are distinguishable in the final visualization, once again proving the usefulness of Pyvis' front-end features.
\end{DUlineblock}

\subsection{Under the Hood%
  \label{under-the-hood}%
}
VisJS reduces the definition of a network to a declarative set of objects. Nodes, Edges, and an Options JSON object are given to the VisJS Network constructor. The following basic example from their documentation proves this:\vspace{1mm}
\begin{Verbatim}[commandchars=\\\{\},fontsize=\footnotesize]
\PY{c+c1}{// create an array with nodes}
\PY{k+kd}{var} \PY{n+nx}{nodes} \PY{o}{=} \PY{k}{new} \PY{n+nx}{vis}\PY{p}{.}\PY{n+nx}{DataSet}\PY{p}{(}\PY{p}{[}
   \PY{p}{\PYZob{}}\PY{n+nx}{id}\PY{o}{:} \PY{l+m+mi}{1}\PY{p}{,} \PY{n+nx}{label}\PY{o}{:} \PY{l+s+s1}{\PYZsq{}Node 1\PYZsq{}}\PY{p}{\PYZcb{}}\PY{p}{,}
   \PY{p}{\PYZob{}}\PY{n+nx}{id}\PY{o}{:} \PY{l+m+mi}{2}\PY{p}{,} \PY{n+nx}{label}\PY{o}{:} \PY{l+s+s1}{\PYZsq{}Node 2\PYZsq{}}\PY{p}{\PYZcb{}}\PY{p}{,}
\PY{p}{]}\PY{p}{)}\PY{p}{;}

\PY{c+c1}{// create an array with edges}
\PY{k+kd}{var} \PY{n+nx}{edges} \PY{o}{=} \PY{k}{new} \PY{n+nx}{vis}\PY{p}{.}\PY{n+nx}{DataSet}\PY{p}{(}\PY{p}{[}
   \PY{p}{\PYZob{}}\PY{n+nx}{from}\PY{o}{:} \PY{l+m+mi}{1}\PY{p}{,} \PY{n+nx}{to}\PY{o}{:} \PY{l+m+mi}{2}\PY{p}{\PYZcb{}}\PY{p}{,}
\PY{p}{]}\PY{p}{)}\PY{p}{;}

\PY{c+c1}{// create a network}
\PY{k+kd}{var} \PY{n+nx}{container} \PY{o}{=} \PY{n+nb}{document}\PY{p}{.}\PY{n+nx}{getElementById}\PY{p}{(}\PY{l+s+s1}{\PYZsq{}mynetwork\PYZsq{}}\PY{p}{)}\PY{p}{;}

\PY{c+c1}{// provide the data in the vis format}
\PY{k+kd}{var} \PY{n+nx}{data} \PY{o}{=} \PY{p}{\PYZob{}}
   \PY{n+nx}{nodes}\PY{o}{:} \PY{n+nx}{nodes}\PY{p}{,}
   \PY{n+nx}{edges}\PY{o}{:} \PY{n+nx}{edges}
\PY{p}{\PYZcb{}}\PY{p}{;}
\PY{k+kd}{var} \PY{n+nx}{options} \PY{o}{=} \PY{p}{\PYZob{}}\PY{p}{\PYZcb{}}\PY{p}{;}

\PY{c+c1}{// initialize your network!}
\PY{k+kd}{var} \PY{n+nx}{network} \PY{o}{=} \PY{k}{new} \PY{n+nx}{vis}\PY{p}{.}\PY{n+nx}{Network}\PY{p}{(}\PY{n+nx}{container}\PY{p}{,} \PY{n+nx}{data}\PY{p}{,} \PY{n+nx}{options}\PY{p}{)}\PY{p}{;}
\end{Verbatim}
\vspace{1mm}

\begin{DUlineblock}{0em}
\item[] This pattern makes Jinja \DUrole{cite}{jinja} templating an obvious candidate for generalizing a set of JavaScript declarations. VisJS documentation provides a complete set of supported attributes for each data structure, so incorporating them into the Python layer involves representing each object as Python objects which are then serialized and sent to Jinja to handle the templating.
\item[] A simple example of this process in action is outlined below:
\end{DUlineblock}
\vspace{1mm}
\begin{Verbatim}[commandchars=\\\{\},fontsize=\footnotesize]
\PY{n+nb+bp}{self}\PY{o}{.}\PY{n}{html} \PY{o}{=} \PY{n}{template}\PY{o}{.}\PY{n}{render}\PY{p}{(}\PY{n}{nodes}\PY{o}{=}\PY{n}{nodes}\PY{p}{,} \PY{n}{edges}\PY{o}{=}\PY{n}{edges}\PY{p}{)}
\end{Verbatim}
\vspace{1mm}

\begin{DUlineblock}{0em}
\item[] In this case, a template HTML file is rendered with node and edge data matching a format compatible with a VisJS Network instance.
\end{DUlineblock}

\subsection{Conclusion%
  \label{conclusion}%
}
Pyvis is a powerful python module for visualizing and interactively manipulating network graphs in a standalone web application or a Jupyter notebook. Pyvis brings the power of VisJS to Python, thus enabling data scientists who use Jupyter to interactively visualize network graphs with all the fluid interactions of a pure-JavaScript application.

Code samples presented here, and with the corresponding poster presentation, as well as other supplemental material are available at West Health's github repository at
\href{https://github.com/Westhealth/scipy2020/pyvis/}{https://github.com/Westhealth/scipy2020/pyvis}.
\bibliographystyle{alphaurl}
\bibliography{ourbib}

\end{document}